\documentclass{kapproc} 

\usepackage[T1]{fontenc}

\usepackage{procps} 

\usepackage[dvips]{graphicx}

\upperandlowercase

\kluwerbib 

\begin{document}



\articletitle[A multi-wavelength study of NGC 7673]{
A multi-wavelength study of \\
the starburst galaxy NGC 7673
}



















\author{Paola Castangia\altaffilmark{1,2}, Anna Pasquali\altaffilmark{3}, 
         Piero Benvenuti\altaffilmark{1}}

\affil{\altaffilmark{1}Universit\`{a} di Cagliari (Italy), \ 
\altaffilmark{2}INAF-Osservatorio Astronomico di Cagliari (Italy), \\
\altaffilmark{3}ETH Z\"{u}rich (Switzerland)}


\begin{abstract}

HST/WFPC2 images resolved NGC~7673 into a large number of star clusters. Among these, 31 fall inside the IUE large aperture, which was used to acquire an integrated ultraviolet spectrum of the galaxy. We have fitted Starburst99 evolutionary synthesis models  to the observed colours of these clusters and derived ages, reddenings, and masses of the clusters. Then a simple sum of the clusters synthetic UV spectra has been  compared to the observed IUE spectrum, in order to investigate the star formation history of NGC~7673.

\end{abstract}



\section{Introduction}

The high angular resolution of the Hubble Space Telescope, has significantly improved our understanding of nearby starburst galaxies. One of the most intriguing results of the HST imaging has been the discovery of compact high-luminosity star clusters (often referred to as "super star cluster", SSCs), in a variety of galactic enviroments. These SSCs show remarkably similar properties: compact sizes (few tens of parsec); young ages (< 1Gyr); and associated high luminosities, implying masses between $10^5$ and $10^6$ M$_\odot$.
So far, several studies indicate that 20\% to 50\% of the recent star formation occurs in clusters, regardless of the location within the galaxy (e.g. galactic disks or circumnuclear rings) and of its Hubble type (from spirals to mergers; \cite{buat,maoz,meurer}).

UV imaging has shown that the morphology of starburst regions is quite complex, with compact, bright, young clusters embedded in a diffuse irregular UV background. Buat et al. (1994) have found that discrete sources dominate the UV emission of M33 and that the fraction of the UV flux arising from diffuse regions is not larger than 20\%. On the contrary,  Maoz et al. (1996) have determined that the fraction of UV flux emitted by star clusters can vary between 15\% and 50\% of the total UV luminosity of circumnuclear rings in five nearby spirals. This result is confirmed by Meurer et al. (1995) who have studied a sample of nine starburst galaxies ranging in morphology from isolated BCDs to FIRGs in merging systems. They have found that the fraction of ultraviolet flux from clusters varies between 5\% and 52\% of the total galaxy flux in UV, with an average of 20\%.
The nature of the diffuse emission is still a matter of debate. In principle it could be cluster light reflected by dust as well as emission arising from unresolved stars, not bounded in clusters. For the very few galaxies which are near enough that individual stars can be resolved, it has been demostrated that the diffuse light comes from the field stellar population (\cite{meurer,tremo}). 

It is clear from the above examples that the percentage of contributed ultraviolet light from clusters varies dramatically over a large range of galaxy morphologies. What constrains the amount of star formation in clusters against the one in the field? Unfortunately the statistics is still very poor and more observations are needed to answer this question. In this paper we present the first results of our analysis of the UV spectrum of NGC~7673. 

NGC~7673 is a fascinating, nearby ($d=49$ Mpc) starburst, whose star formation activity is concentrated in large complexes, the distinctive "clumps", clearly visible in the optical images as several bright knots in the galactic disk. Since initial investigations, these clumps were found to be hyperactive star-forming regions. Benvenuti et al. (1982) discovered that they radiate in the far UV, on average, 100 times more than the giant {\hbox{H{\sc ii}}} region 30 Doradus. Recently, these clumps have been resolved by HST/WFPC2 images into a large number of star clusters (\cite{hgp}). Which process might have triggered the starburst activity is not yet well understood. Despite the disturbed optical appearance (the spiral pattern is asymmetric) and the large-scale starburst, H$_\alpha$ kinematics (\cite{hg}) demonstrates that NGC~7673 is a relatively unperturbed, rotating disk galaxy seen nearly face-on. Thus an ongoing interaction of NGC~7673 with its neighbour NGC~7677 can be ruled out as the trigger of the recent star formation in NGC~7673, though a past interaction or a minor merger with a dwarf companion cannot be excluded (\cite{hg}).
31 of the star clusters in NGC~7673, fall inside the IUE large aperture which was used to acquire an integrated UV spectrum of the galaxy. We have fitted Starburst99 evolutionary models to the observed optical colours of these clusters and derived ages, reddenings, masses, and UV spectra. Then a simple sum of these synthetic spectra has been compared to the observed IUE spectrum in order to estimate the contribution of the star clusters to the UV luminosity of the galaxy. 

\section{The UV spectrum}

\begin{figure}[htbp]
\centering
\includegraphics[width=0.9\textwidth]{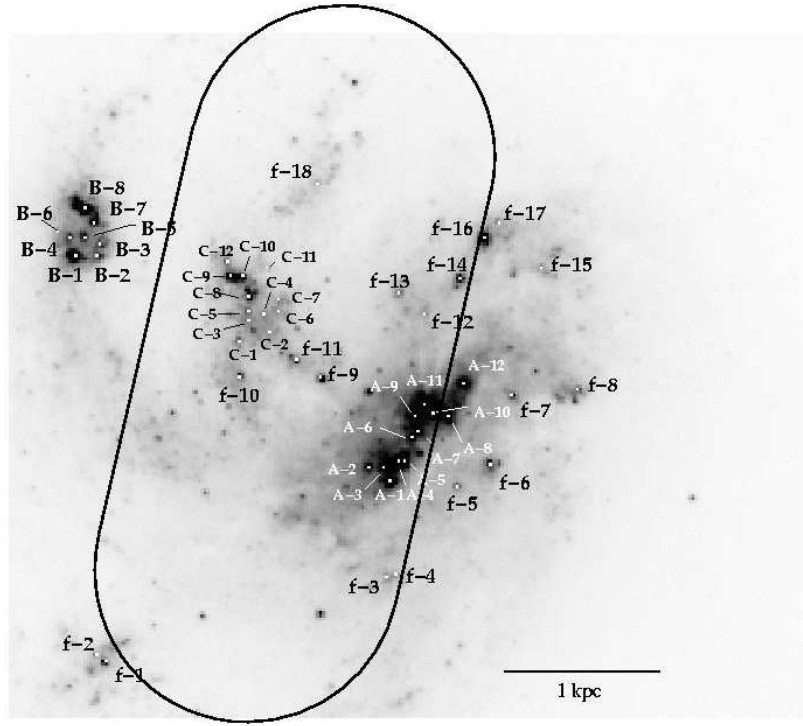}
\caption{Clusters positions superimposed on the HST/WFPC2 image obtained with the F555W filter (\cite{hgp}). The position and size of the IUE slit is also indicated.}
\label{mappa}
\end{figure}

The observed UV spectrum of NGC~7673, was obtained with the International Ultraviolet Explorer (IUE) in 1979, in the short wavelength range 1150-2000 \AA, and in the low dispersion mode (the resolution is about 6 \AA/pix). The large slit $10''\times 20''$ was used to acquire the integrated spectrum (see Fig. \ref{mappa}). Before comparing it to the synthetic spectra, we  corrected the IUE spectrum for galactic reddening, using Seaton's (1979) extinction law and shifted it to restframe wavelengths.

\section{The clusters parameters}

\begin{table}[p]
\rotatebox{90}{\vbox to\textwidth{
\hsize=\textheight
\caption{Columns 2-4 list the observed apparent magnitude and colours of the clusters which fall inside the IUE slit (\cite{hgp}). Columns 5-13 are the best-fitting ages, reddenings, and masses, together with their minimum and maximum value.}
\begin{tabular*}{\hsize}{@{\extracolsep{\fill}}*{13}{c}}
\sphline
\it & & & & & \it Minimum & \it Maximum & & \it Minimum & \it Maximum & & \it Minimum & \it Maximum \cr
\it Cluster & \it V & \it U-V & \it V-I & \it Age & \it age & \it age & \it E(B-V) & \it E(B-V) & \it E(B-V) & \it Mass & \it mass & \it mass\cr
\it & \it (mag) & \it (mag) & \it (mag) & \it (Myr) & \it (Myr) & \it (Myr) & \it (mag) & \it (mag) & \it (mag) & \it (10$^6$ M$_\odot$) & \it (10$^6$ M$_\odot$) & \it (10$^6$ M$_\odot$)\cr
\sphline
A-1 & 19.80 & -1.09 & 0.35 & 4.01 & 4.01 & 22.01 & 0.57 & 0.10 & 0.61 & 1.3 & 1.1 & 2.0\cr
A-2 & 20.38 & -0.19 & 0.53 & 16.01 & 6.01 & 308.01 & 0.44 & 0.00 & 0.78 & 1.3 & 0.6 & 4.5\cr
A-3 & 19.53 & -1.36 & 0.55 & 10.01 & 0.01 & 28.01 & 0.00 & 0.00 & 0.89 & 0.5 & 0.3 & 19.8\cr
A-4 & 19.94 & -1.38 & 0.02 & 6.01 & 6.01 & 6.01 & 0.00 & 0.00 & 0.01 & 0.2 & 0.2 & 0.2\cr
A-5 & 19.91 & -1.63 & 0.11 & 4.01 & 0.01 & 4.01 & 0.22 & 0.16 & 0.61 & 0.3 & 0.3 & 4.6\cr
A-6 & 20.63 & -0.79 & 0.65 & 28.01 & 0.01 & 74.01 & 0.26 & 0.00 & 1.25 & 1.3 & 0.2 & 25.8\cr
A-7 & 20.66 & -1.12 & -0.06 & 6.01 & 6.01 & 6.01 & 0.08 & 0.06 & 0.10 & 0.1 & 0.1 & 0.1\cr
A-9 & 19.97 & -0.96 & 0.43  & 18.01 & 4.01 & 46.01 & 0.32 & 0.00 & 0.70 & 2.0 & 1.0 & 2.4\cr
A-10 & 20.37 & -1.11 & 0.46 & 8.01 & 0.01 & 38.01 & 0.09 & 0.00 & 1.00 & 0.2 & 0.1 & 12.9\cr
A-11 & 19.07 & -0.08 & 0.61 & 16.01 & 4.01 & 454.01 & 0.60 & 0.00 & 1.27 & 7.7 & 2.4 & 35.7\cr
C-1 & 21.29 & -1.64 & 0.10 & 4.01 & 0.01 & 4.01 & 0.22 & 0.16 & 0.56 & 0.1 & 0.1 & 1.0\cr
C-2 & 22.22 & -1.80 & 0.24 & 2.01 & 0.01 & 4.01 & 0.51 & 0.06 & 0.68 & 0.3 & 0.0 & 0.8\cr
C-3 & 21.80 & -1.58 & 0.43 & 2.01 & 0.01 & 2.01 & 0.68 & 0.57 & 0.78 & 0.9 & 0.6 & 1.6\cr
C-5 & 21.55 & -1.71 & 0.36 & 2.01 & 0.01 & 2.01 & 0.58 & 0.51 & 0.64 & 0.7 & 0.6 & 1.2\cr
C-6 & 21.90 & -1.85 & 0.63 & 2.01 & 0.01 & 48.01 & 0.56 & 0.00 & 1.09 & 0.6 & 0.0 & 6.2\cr
C-7 & 21.24 & -1.63 & 0.39 & 2.01 & 0.01 & 38.01 & 0.76 & 0.00 & 0.96 & 2.0 & 0.1 & 5.8\cr
C-8 & 20.14 & -1.71 & 0.36 & 2.01 & 0.01 & 20.01 & 0.58 & 0.00 & 0.78 & 2.7 & 0.2 & 7.7\cr
C-9 & 19.83 & -1.38 & 0.07 & 6.01 & 4.01 & 6.01 & 0.00 & 0.00 & 0.35 & 0.2 & 0.2 & 0.6\cr
C-10 & 19.89 & -1.84 & 0.19 & 2.01 & 0.01 & 2.01 & 0.48 & 0.37 & 0.59 & 2.3 & 1.5 & 4.5\cr
C-11 & 22.32 & -2.06 & 0.07 & 2.01 & 0.01 & 2.01 & 0.34 & 0.28 & 0.40 & 0.2 & 0.1 & 0.2\cr
C-12 & 21.01 & -1.25 & 0.42 & 2.01 & 0.01 & 28.01 & 0.84 & 0.00 & 0.87 & 3.0 & 0.1 & 4.6\cr
F-3 & 21.73 & -1.55 & 0.25 & 2.01 & 0.01 & 20.01 & 0.63 & 0.00 & 0.70 & 0.7 & 0.1 & 1.2\cr
F-4 & 21.74 & -1.38 & 0.32 & 0.01 & 0.01 & 22.01 & 0.72 & 0.00 & 0.81 & 1.4 & 0.1 & 1.8\cr
F-9 & 20.41 & -0.98 & 0.65 & 10.01 & 10.01 & 26.01 & 0.19 & 0.04 & 0.22 & 0.4 & 0.3 & 1.3\cr
F-10 & 21.33 & -1.91 & 0.30 & 2.01 & 0.01 & 20.01 & 0.49 & 0.00 & 0.76 & 0.7 & 0.1 & 2.6\cr
F-11 & 20.96 & -1.23 & 0.59 & 12.01 & 10.01 & 12.01 & 0.00 & 0.00 & 0.08 & 0.2 & 0.1 & 0.2\cr
F-12 & 22.02 & -1.73 & 0.24 & 2.01 & 0.01 & 2.01 & 0.56 & 0.50 & 0.60 & 0.4 & 0.4 & 0.7\cr
F-13 & 21.33 & -1.91 & 0.30 & 2.01 & 0.01 & 20.01 & 0.48 & 0.00 & 0.74 & 0.6 & 0.1 & 2.3\cr
F-14 & 20.27 & -2.01 & -0.19 & 2.01 & 0.01 & 4.01 & 0.42 & 0.00 & 0.63 & 1.3 & 0.1 & 4.1\cr
F-18 & 22.01 & -1.52 & 0.30 & 2.01 & 0.01 & 20.01 & 0.68 & 0.00 & 0.76 & 0.7 & 0.1 & 1.2\cr
\sphline
\end{tabular*}
}}
\end{table}

The observed (U - V) and (V -  I) colours of each cluster  were corrected for galactic extinction and fitted with a Starburst99  model, computed for a total cluster mass of $10^6$ M$_\odot$ and LMC-like metallicity. 
The fitting technique, based on minimazing the fit $\chi^2$, is described in detail in Pasquali et al. (2003). Here, it is worthwhile to recall that the clusters mass is derived by scaling the mass of the Starburst99 model by the ratio of the dereddened apparent magnitude to the one of the best-fitting age-reddening solution. 
Since only two colours are available, our fitting procedure does not yield unique solutions. Therefore, together with the age, reddening, and mass computation corresponding to the minimum $\chi^2$, we have also selected all the solutions with $\chi^2 < 1.5 {\chi^2}_{min}$. The latters define a minimum and a maximum value (for age, reddening, and mass) within which the cluster parameters are considered equally consistent with the data and also define the uncertainties on the best-fitting parameters.

\section{Results}

\begin{figure}[htbp]
\centering
\includegraphics[width=\textwidth]{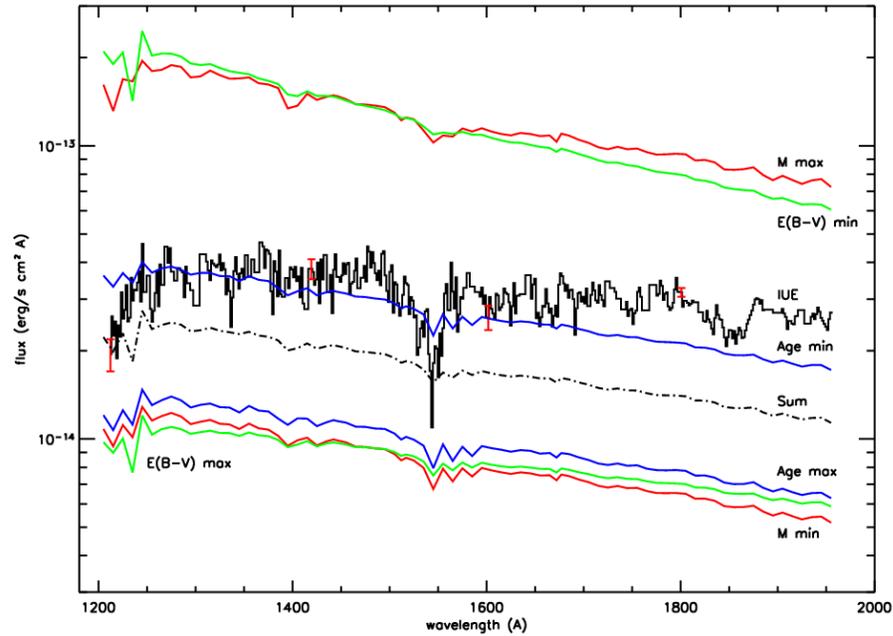}
\caption{Synthetic spectra superimposed on the observed IUE spectrum of NGC~7673. The dot-dashed line represents the synthetic spectrum corresponding to the best fitting parameters. Blue, green, and red lines correspond to the minimum  and maximum ages, reddenings, and masses, respectively.}
\label{spettri}
\end{figure} 

In order to compare the IUE spectrum of NGC~7673 with the simple sum of the clusters synthetic spectra, we have reddened each cluster spectrum by the E(B-V) listed in Table 1 using Calzetti's (2001) extintion curve and scaled it to the distance of NGC~7673. Specifically, for each cluster we have synthetized the spectrum  corresponding to the best fitting parameters, and a set of spectra corresponding to the minimum and maximum values of the cluster parameters. For each choice of parameters we have simply summed the clusters spectra together. Fig. \ref{spettri} shows the observed IUE spectrum of NGC~7673, plotted together with: the sum of the spectra obtained for the clusters best fitting parameters (dot-dashed line); the sums corresponding to the cluster minimum and maximum ages (blue solid lines); the sums corresponding to the cluster minimum and maximum reddenings (green solid lines); the sums corresponding to the cluster minimum and maximum masses (red solid lines).

Blueward of 1600 \AA, the theoretical flux computed from the clusters best fitting parameters is about 65\% of the observed, and it increases to almost 100\% when calculated in correspondance of the clusters youngest ages (and assuming the best fitting reddenings and masses). At longer wavelengths these two fractions decrease to 50\% and 70\%, respectively. We believe that the decrease with wavelength of the contributed flux from star clusters may be due to the adopted reddening and reddening law. This effect will be further investigated (Castangia et al. 2005, in preparation). 

At face value, the above results are somewhat in contrast with Homeier et al. (2002), who estimated that only between 16\% and 33\% of the luminosity of the clumps in the F255W filter is due to the star clusters. According to our estimates, the fraction of UV flux emitted by clusters in NGC~7673 is generally larger than in any other galaxy in which this kind of measure has been performed (\cite{meurer,maoz}), with the only exception of M33 (\cite{buat}). 
On the other hand, we would agree with the results of Homeier et al., Meurer et al., and Maoz et al. only when we consider the sums of the synthetic spectra computed for the clusters oldest ages, highest reddenings, and lowest masses. 
Indeed these spectra give a flux contribution from star clusters to the galaxy UV emission of about 30\%.






%




\end{document}